# Is AI Changing the Rules of Academic Misconduct? An In-depth Look at Students' Perceptions of '*AI-giarism*'

Cecilia Ka Yuk Chan

Professor, Faculty of Education,

The University of Hong Kong

Email: Cecilia.Chan@cetl.hku.hk

**Abstract**

This pioneering study explores students' perceptions of *AI-giarism*, an emergent form of academic dishonesty involving AI and plagiarism, within the higher education context. A survey, undertaken by 393 undergraduate and postgraduate students from a variety of disciplines, investigated their perceptions of diverse AI-giarism scenarios. The findings portray a complex landscape of understanding, with clear disapproval for direct AI content generation, yet more ambivalent attitudes towards subtler uses of AI. The study introduces a novel instrument, as an initial conceptualization of AI-giarism, offering a significant tool for educators and policy-makers. This scale facilitates understanding and discussions around AI-related academic misconduct, aiding in pedagogical design and assessment in an era of AI integration. Moreover, it challenges traditional definitions of academic misconduct, emphasizing the need to adapt in response to evolving AI technology. Despite limitations, such as the rapidly changing nature of AI and the use of convenience sampling, the study provides pivotal insights for academia, policy-making, and the broader integration of AI technology in education.

**1. Introduction: AI and Human or AI Vs Human**

Over the past few months, academic integrity has been under open debate (Chan & Hu, 2023; Gendron, Andrew & Cooper, 2022; Liebrenz et al, 2023; Stokel-Walker, 2023), driven by the growing popularity of the text-based generative AI platform ChatGPT (OpenAI, 2023). Adhering to the ethical considerations in academic writing is critical for ensuring fairness, integrity, authenticity, safeguarding the rights of research, maintaining transparency and accuracy, and upholding academic standards (Chan & Lee, 2023; Hosseini, Rasmussen, & Resnik, 2023). Some of the key academic integrity principles for writing and research include authorship criteria, disclosure of conflict of interest, ensuring data integrity and accuracy, avoiding plagiarism and research misconduct, respect for intellectual property, and being willing to retract or correct published articles if errors or misconduct are discovered. It is the ground zero for monitoring the quality of academic integrity for scientific research and students' academic work.

However, the rapid advancement of generative AI tools presents a serious threat to academic integrity. The primary issue revolves around the misuse of AI tools by students and researchers, who may use these AI capabilities to produce essays, research reports, or even entire theses, presenting the AI-generated work as their own (Chan & Tsi, 2023; Cotton, Cotton, & Shipway, 2023). This misuse could shift the focus of the learning process from acquiring, applying and critiquing knowledge to merely generating the output that may be factually incorrect. This process undermines the core principles of academic integrity, promotes academic dishonesty, and detracts students from their learning journey towards creativity, originality, and critical thinking.

Another significant concern is the difficulty in differentiating between human-authored text and AI-generated text. As the capabilities of generative AI models such as GPT-4 improve, the produced content becomes increasingly coherent and contextually relevant, making it challenging to distinguish between AI-generated text and human-written text. This indistinguishability complicates the efforts of educators to assess the authenticity of a student's work or a researcher's findings, thereby posing a problem for upholding academic integrity.

While AI software can effectively detect traditional forms of plagiarism, they may fail to identify instances where AI tools have been used to produce the work. Current plagiarism detection tools are generally designed to identify text that matches entries in a database, but they may not be equipped to detect AI-generated content that is 'original' in terms of its wording but not in terms of its ideas or structure. This limitation in detection could lead to an increase in undetected instances of academic dishonesty, further compromising academic integrity (Ahmad, Murugesan, & Kshetri, 2023).

While the launch of ChatGPT has brought generative AI to the forefront of academic integrity discussions, it's important to be mindful that AI has been working with different industries side by side for years, including medical innovation, drug discovery, computer imaging and research, and education administration, with machines co-working with humans on these applications (Jiang et al., 2017). As we traverse in this AI-dominated era, it appears that the integration of AI and human interaction is an unceasing phenomenon, perhaps the idea of human machine co-writing should be the norm. Given this trend, it raises the question of what constitutes unethical behaviour in academic writing including plagiarism, attribution, copyrights, and authorship in the context of AI-generated content. And in this study, we will attempt to answer this question by gathering students' current understanding of plagiarism and AI-generated writing.

## 2. Conceptualisation of Plagiarism

Plagiarism is the act of using someone else's work without proper acknowledgement or permission, passing it off as one's own original work. Park (2003) has defined it as "literary theft," while Freedman (1994, p. 517) expressed a similar sentiment, believing plagiarism is "an attack on individuality, on nothing less than a basic human right." Most universities have a zero-tolerance rule on plagiarism (University of Hong Kong, n.d.). The left-hand side of figure 1 shows the common zero-tolerance rules on plagiarism found among most universities, from one end of the scale "completed the work totally by oneself, using proper citations where necessary" to the other end "using exact words from a source without due acknowledgement of the source". However, researchers and practitioners have noted that the act of plagiarism is complicated and ambiguous "between imitation and theft, between borrowing and plagiarism, lies a wide, murky borderland" (as cited in Park, 2003, p. 475). In some cases, authors may not even be aware of the original sources of the content, and may treat it as their own work without proper attribution; as Leatherman (1999) states "the point at which an idea passes into general knowledge in a way that no longer requires attribution". Wagner (2014) discusses the ongoing issue of plagiarism in scholarly works and the various factors that influence the definition and handling of plagiarism. Factors such as the extent of copying, originality, positioning, referencing, the intention of the authors, their seniority, and the language they are writing in are all important considerations when defining plagiarism (Wagner, 2014). The extent of plagiarism can range from copying a few sentences to entire papers or chapters, the latter of which also breaches copyright laws. The position of the plagiarised material within the work, whether it's properly referenced, the author's intention behind plagiarising, and the seniority and native language of the author are all factors that can influence the severity of the plagiarism (Wagner, 2014). Wagner's article also notes that while summarizing other works is common in scholarly writing, it can be difficult to distinguish between acceptable summarising and outright copying, and thus, highlights the need for clearer definitions and guidelines for plagiarism.

As mentioned above, one of the most pressing issues that the advancement in AI technologies brings to the educational context is the threat to academic integrity. Educators are particularly concerned about the potential misuse of AI to evade plagiarism detection. Before delving deeper into the issue, it is important to review some of the notable findings from research on investigating perceptions of plagiarism from the perspectives of students.

<< Figure 1: AI-giarism and Plagiarism Literacy for academic writing – what constitutes unethical?>>

## 3. Student Perception of Plagiarism

Over the years, numerous studies have aimed to assess students' understanding of plagiarism and their ability to avoid it. Some scholars have taken a step further by attempting to develop validated psychometric instruments. These tools provide a more reliable measurement of students' perceptions of plagiarism. For instance, Mavrinac et al. (2010) created an instrument to measure students' positive attitudes, negative attitudes, and subjective norms towards plagiarism. Subsequently, Howard, Ehrich, & Walton (2014) revised Mavrinac et al. (2010)'s instrument for further refinement. Oghabi, Pourdana, & Ghaemi (2020) contributed to this body of research by developing a Sociocultural Plagiarism Questionnaire. Their questionnaire evaluates students' theoretical and practical understanding of plagiarism. Furthermore, Cheung, Stupple, & Elander (2017) developed a psychometric measure focusing on students' attitudes and beliefs about authorship as a means to prevent plagiarism.

Apart from these studies, there are others that used validated surveys and questionnaires to analyse students' perceptions of plagiarism. They revealed different levels of awareness and knowledge about plagiarism across various countries and disciplines. For instance, studies on Australian students from the Global North demonstrated a high level of awareness and understanding of plagiarism (Bretag et al., 2014; Smedley, Crawford & Cloete, 2015). However, research on Croatian (Bašić et al., 2019) and Canadian students (Bokosmaty et al., 2019) unveiled troubling results. These students appeared to struggle with adhering to referencing rules and avoiding self-plagiarism.

In contrast, research conducted in countries of the Global South highlighted several factors contributing to an insufficient awareness and understanding of plagiarism. These factors include language barriers that encourage a more lenient attitude towards plagiarism (Erguvan, 2022), confusion about referencing rules (Hussein, 2022; Ibegbulam & Eze, 2015), cultural differences that foster tolerance for plagiarism behaviors, such as the emphasis on personal relations over copyright rules, and an entrenched belief in the merits of memorization and imitation (Hu & Lei, 2012; Oghabi, Pourdana & Ghaemi, 2020). There are also misconceptions about contract cheating and the practice of paying others to substitute for essay writing (Romanowski, 2022).

Given the incomplete understanding of plagiarism within higher education, several scholars have emphasized in their research the need for further guidance, policies, and training workshops to enhance students' research writing and academic referencing skills (Issrani et al., 2021; Rathore et al., 2015).

## 4. Introduction of AI-giarism

Discussions continue among scholars and practitioners about defining and conceptualizing plagiarism. Simultaneously, the use of AI in academic writing has generated significant interest in AI related plagiarism. This has led to the emergence of a new term that warrants our attention: AI-giarism, a term that combines 'AI' and 'plagiarism', and has yet to be widely researched and defined within academic literature. For the study of AI in education in relation to academic misconduct, I, propose the following definition:

*AI-giarism refers to the unethical practice of using artificial intelligence technology, particularly generative language models, to generate content that is plagiarised either from original human-authored work or directly from AI-generated content, without appropriate acknowledgement of the original sources or AI's contribution.*

This can happen when people use AI tools or language models to create written or multimedia content, such as articles, blog posts, images or videos, without properly attributing the original sources or modifying the content sufficiently to make it original (Salvagno et al, 2023). Some AI tools are designed to generate content by automatically combining or predicting from different sources, such as articles or websites, and paraphrasing them to create new content. AI-giarism blurs traditional boundaries of authorship and plagiarism, raising new questions about academic integrity in the digital age.

## 5. Current APA and MLA citation guidelines for AI generated content

Many publishers follow the American Psychological Association (APA) and Modern Language Association (MLA) citation guidelines, and thus it is important to mention their current recommendations for AI generated content. Both APA and MLA suggest when referencing a language model like ChatGPT, it is important to provide detailed information about the specific version of the model, training data, and the organization responsible for its development. In-text citations should include the organization's name (e.g., OpenAI) and the publication year. The reference list entry should include the title of the language model (e.g., ChatGPT), the version number, the organization responsible for the model (e.g., OpenAI), and the URL where the model can be accessed. Furthermore, for enhanced transparency and accessibility, APA (McAdoo, 2023) advises to incorporate the complete text of extended responses generated by ChatGPT as an appendix within the paper or as online supplemental materials, allowing readers to examine the precise text that was produced. MLA (MLA, n.d.) recommends incorporating a descriptive title for the source when citing content generated by an AI tool. This process may entail providing information about the prompt within the title of source element, particularly if it has not been mentioned in the text. Such a practice ensures a comprehensive and accurate representation of the AI-generated content in the citation. For example, to cite the below

|  | In 150 words, explain the theme in Charlie and the Chocolate Factory by Roald Dahl |
|---|---|
| 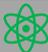 | Charlie and the Chocolate Factory, a renowned children's novel by Roald Dahl, delves into themes such as morality, social justice, imagination, and the consequences of indulgence. The story follows |

> Charlie Bucket, a humble boy from a poor family who, along with four other children, wins a tour of Willy Wonka's enigmatic chocolate factory. The children's contrasting personalities highlight the theme of morality, as Charlie's kindness and selflessness triumph over the other children's vices, which lead to their downfalls. The novel also explores social justice through Charlie's economic struggles and ultimate inheritance of the factory, emphasizing that moral values can prevail over materialism. Imagination is celebrated through the fantastical world of Wonka's factory, encouraging readers to embrace creativity. Finally, the story underscores the dangers of indulgence, as the flawed children face misfortunes due to their inability to resist temptation, while Charlie's self-restraint leads to his reward.

Quoted in the article

> When we study social justice in society, a classic children's novel written by Roald Dahl, Charlie and the Chocolate Factory is often emphasised. It "explores social justice through Charlie's (the protagonist) economic struggles and ultimate inheritance of the factory, emphasizing that moral values can prevail over materialism" ("In 150 words").

Works-Cited-List Entry

> "In 150 words, explain the theme in Charlie and the Chocolate Factory by Roald Dahl" follow-up prompt to list sources. ChatGPT, Mar 23 version, OpenAI, 18 April 2023, chat.openai.com/chat.

Referencing AI-generated content can be a complex and nuanced task, as there are many different types of AI models and applications, each with their own unique characteristics and data sources.

## 6. The Study

It is important to develop clear and consistent guidelines and policy for referencing AI-generated content (Chan, 2023; Sabzalieva & Valentini, 2023) in order to ensure that researchers and students are able to provide proper attribution and credit for the sources they use in their work. These guidelines should be informed by stakeholders, ethical considerations and maybe even, legal compliance, and should take into account the potential implications of using AI-generated content in research. However, there are many considerations. One challenge in referencing AI-generated content is that the same model or dataset can be used by multiple researchers to generate different outputs, which can make it difficult to trace the origin of specific pieces of content. Reproducing the exact wording is often not feasible. As a solution, it may be essential to devise standardized citation methods for the particular version of the AI model or dataset employed, in addition to the specific parameters or settings utilized in content generation as shown in the MLA and APA sections above (McAdoo, 2023; MLA, n.d.). Naturally, this raises the question of how much evidence researchers should provide for each citation.

Another important consideration is the potential ethical implications of using AI-generated content in research, particularly if the content contains sensitive or personal information. Researchers should take steps to ensure that the content is used in a responsible and ethical manner, and that it is not used to harm or discriminate against individuals or groups.

Other consideration lies on the actual conceptualisation of AI-giarism. Generative text AI can provide ideas based on a single text prompt input, if the initial prompt was originated by the author, does that count as AI-giarism? The right-hand side of figure 1 illustrates two extremes of AI-giarism: Clear-cut AI-giarism at one end and Human written content at the other. However, is there an in-between or a grey area where AI-generated content may be considered borderline (ambivalent) AI-giarism as opposed to zero tolerance AI-generated plagiarism?

In this study, we explore students' perception of adopting generative AI for research and study purposes, and investigate their understanding of traditional plagiarism and their perception of AI-plagiarism. It is critical to understand both students' perceptions of traditional plagiarism and their understanding of AI-giarism within the same study, for several reasons.

Firstly, it provides a point of reference. Traditional plagiarism is a well-established concept within academia. It has a comprehensive body of literature that outlines what constitutes plagiarism and the potential consequences for those involved. By analysing students' understanding of traditional plagiarism, the study provides a reference point from which to understand students' comprehension of AI-giarism.

Secondly, it helps identify gaps in understanding or knowledge. By examining students' perceptions of traditional plagiarism and AI-giarism in tandem, it allows for a comparison between the two. This comparison can highlight areas where students' understanding of AI-giarism may fall short or differ significantly from their understanding of traditional plagiarism, offering insights into where educational initiatives might be directed to close those gaps.

Furthermore, understanding students' perceptions of both concepts is integral to evolving academic integrity policies. With the rise of AI technologies and their increasing incorporation into higher education, there is an emerging need to consider how the conventional boundaries of academic integrity apply in these new contexts. A detailed understanding of students' perceptions of both traditional plagiarism and AI-giarism is crucial in shaping relevant, robust, and comprehensive academic integrity policies that appropriately address the evolving landscape of higher education.

Finally, exploring the understanding of traditional plagiarism and AI-giarism together may also help to shed light on broader questions about academic integrity in the digital age. The ways in which students perceive and differentiate between these two forms of academic misconduct can provide insights into their general attitudes towards academic integrity, their conceptualization of original work, and their understanding of the ethical implications of using AI technologies in academia.

The integration of both traditional plagiarism and AI-giarism into the same study offers a rich, nuanced picture of students' understanding of academic integrity in the context of rapid technological advancements. It provides a foundation for addressing these challenges effectively and appropriately, with implications for education, policy, and practice in higher education.

The research questions for this study are:

1. From the perspective of students, what is considered a violation of academic integrity when using AI-generated content in higher education?
2. From the perspective of students, to what extent do they understand traditional forms of plagiarism in higher education?

## 7. Methodology

This investigation utilized an online questionnaire to explore the understanding, attitudes, and experiences related to plagiarism and AI-giarism among students in higher education in Hong Kong. The questionnaire contains three parts, the first part includes basic demographics such as gender, name of university, discipline major, year of study, age and their indication of how often out of a five-point Likert scale (from *Never* to *Always*) do they use generative AI technologies like ChatGPT. The second and third parts contain closed questions on traditional plagiarism and AI-plagiarism. The closed questions pertaining to traditional plagiarism are ranked in a five-point Likert scale (from *Strongly disagree* to *Strongly agree*) and were formulated based on the existing plagiarism policy of a university in Hong Kong.

However, for the AI-plagiarism part of the questionnaire, given the emergent nature of AI-giarism as a research area, there were no existing questionnaires or validated measures to draw from. Hence, the author developed the survey questions concerning AI-giarism based on a comprehensive review of relevant literature in the fields of AI in education (Chan, 2023; Crompton & Burke, 2023; Kong et al, 2021; Kumar, 2023; Ng et al, 2021) and policy formation, as well as a careful study of discussions from university forums. This enabled the creation of an original set of questions in a five-point Likert scale (from *Strongly disagree* to *Strongly agree*) that reflect current debates and concerns surrounding the use of AI technologies in academic contexts.

Participants were selected using a convenience sampling technique, with invitations to participate sent via mass emails. Before accessing the survey, respondents were presented with an informed consent form on the online platform, ensuring their understanding of the study's objectives and their rights as participants.

The data gathered were analysed, descriptive analysis was performed on the quantitative data derived from the closed questions to shed light on the participants' experiences, perceptions, and comprehension of plagiarism and AI-giarism in academia.

## 8. Findings

This study involved a total of 393 students from different universities in Hong Kong, almost evenly split between genders with 198 males and 195 females. The average age of the students was approximately 22 years, with a standard deviation of 2.59 years, indicating that the ages ranged from 17 to 28 years old.

When it comes to their fields of study, the most prevalent discipline was Engineering, with 112 students. Other disciplines included Business 54, Science 58, Arts 61, Architecture 30, Education 30, Social Sciences 26, Law 5, Medicine and Dentistry 12. There were also 5 students who did not specify their discipline.

In terms of the level of study, the largest group was undergraduate students, with a total of 244 participants. Among these, the distribution across the years of study was relatively even: there were 81 freshmen, 51 sophomores, 48 juniors, and 58 seniors. A smaller group of six were in their fifth or sixth year of study. The rest of the participants comprised of 111 taught postgraduates and 44 research postgraduates. Out of 393 students surveyed, the average response to the statement "I have used generative AI technologies (GenAI) like ChatGPT" was 2.27, with a standard deviation of 1.65, suggesting that most students have relatively little experience with using such technologies. Table 1 shows the descriptive analysis of the findings.

| **Items (Traditional Plagiarism) To what extend do you agree or disagree with the following actions as forms of plagiarism:** | n | M(SD) | Mdn |
|---|---|---|---|
| E1. Copying word for word from a source without due acknowledgement of the source. | 390 | 4.13 (1.197) | 5.00 |
| E2. Closely paraphrasing, or substantial copying with minor modifications (such as changing grammar, adding a few words or reversing active/passive voices), without due acknowledgement of the source. | 393 | 3.95 (1.122) | 4.00 |
| E3. Translating a source in one language into another language and using it as your own, without due acknowledgement of the source. | 393 | 3.93 (1.177) | 4.00 |
| E4. Collusion or unauthorized collaboration between students on a piece of work without acknowledging the assistance received. | 384 | 3.79 (1.175) | 4.00 |
| E5. Use of the work of another student or third party (e.g. an essay writing service) for submission as one's original work. | 393 | 4.00 (1.194) | 4.00 |
| E6. Submitting part or all of the same assignment for different courses without acknowledging it. | 385 | 3.59 (1.232) | 4.00 |
| E7. Getting a ghostwriter to write your assignment. | 388 | 3.99 (1.267) | 4.00 |
| **Items (AI-giarism) To what extent do you agree or disagree with the following actions as forms of academic misconduct:** | | | |
| F1. The student input a prompt into an AI system, copied the generated response, and submitted it to the teacher. The student acknowledged the use of AI tools. | 386 | 3.44 (1.177) | 4.00 |
| F2. The student employed AI technologies to paraphrase some of the texts for their assignment from other sources without acknowledgement. However, the student acknowledged the use of AI tools. | 384 | 3.37 (1.126) | 4.00 |
| F3. The student input a prompt into an AI system. After verifying all the facts, making edits, adding references, and formatting the generated response, they submitted it to the teacher. The student acknowledged the use of AI tools. | 389 | 2.75 (1.190) | 3.00 |
| F4. The student used their own ideas to generate prompts with inputs from AI technologies to produce multiple AI responses for their assignment. After verifying all the facts, the student then used the best parts, made edits, and submitted the work. The student acknowledged the use of AI tools. | 386 | 2.61 (1.184) | 2.00 |
| F5. The student employed AI technologies to assist in generating initial ideas, and then supplemented them with their own ideas. After verifying all the facts, the student submitted the assignment with parts generated by AI tools and parts written by the student. The student acknowledged the use of AI tools. | 385 | 2.65 (1.214) | 2.00 |
| F6. The student employed AI technologies to assist in generating initial ideas, and then supplemented them with their own ideas. After verifying all the facts, the student rewrote most sections and submitted the assignment. The student acknowledged the use of AI tools. | 387 | 2.55 (1.232) | 2.00 |
| F7. The student employed AI technologies to rephrase some of their own written content for their assignment in order to improve the writing quality. The student acknowledged the use of AI tools. | 388 | 2.35 (1.206) | 2.00 |
| F8. The student drafted their assignment, sought feedback from AI technologies, and made improvements based on the suggestions provided, which may have entailed modifications to grammar and sentence structure. The student acknowledged the use of AI tools. | 393 | 2.30 (1.222) | 2.00 |
| F9. The student employed AI technologies only to assist with the checking of grammar for their assignment. | 392 | 2.09 (1.285) | 2.00 |
| F10. The student employed AI technologies and the Internet as search engines for resources to assist in completing their assignment, but did not incorporate any text directly from these resources in the assignment. | 384 | 2.30 (1.311) | 2.00 |
| F11. The student completed their assignment without using any AI technologies or referring to the Internet. | 385 | 2.01 (1.276) | 1.00 |

Table 1: Descriptive Statistics Students' understanding and perception of plagiarism and AI-giarism

## 8.1 Students' Conceptualisation of Plagiarism

The students were asked to rate their level of agreement with various forms of plagiarism, based on a 5-point Likert scale. There was a total of 7 items assessed in the questionnaire. The student participants (n ranged from 384 to 393 across items) generally agreed that the listed behaviours are indeed forms of plagiarism, with mean ratings ranging from 3.59 to 4.13.

The lowest agreement was observed for item E6: "Submitting part or all of the same assignment for different courses without acknowledging it." ($M = 3.59$, $SD = 1.232$). This behaviour seemed to be perceived as less serious or perhaps less clearly defined as a form of plagiarism, compared to the other items.

The strongest agreement was for E1: "Copying word for word from a source without due acknowledgement of the source." ($M = 4.13$, $SD = 1.197$), indicating that students most unanimously identified this act as a clear form of plagiarism.

It is noteworthy that the standard deviations for all items were over 1, suggesting a moderate degree of disagreement or diversity in perceptions among the students for each item. Such variation could be related to individual differences in students' understanding of plagiarism or their beliefs about what actions constitute a breach of academic integrity.

Overall, these results highlight students' general understanding of traditional forms of plagiarism in the academic context, yet also underscore some discrepancies in how these behaviours are perceived, pointing to the need for more consistent education on the topic. It would be intriguing to further examine how these perceptions of traditional plagiarism extend to the newer concept of AI-generated plagiarism or 'AI-giarism'.

## 8.2 Students' Perception of AI-giarism

The data represent students' perceptions of different scenarios involving the use of AI in academic writing, on a scale from 1 to 5, where 1 represents strong disagreement and 5 strong agreement that the scenario represents academic misconduct. In other words, a higher score reflects a stronger perception of the scenario as being an instance of academic misconduct, or AI-giarism.

Scenarios F1 and F2, which represent the most direct use of AI tools to generate content for an assignment, are perceived as the most significant instances of academic misconduct among the students, with mean scores of 3.44 and 3.37, respectively. This indicates that students generally understand and respect the principles of academic integrity when it comes to the outright use of AI to generate or paraphrase content.

Scenarios F3 to F7, which describe more nuanced uses of AI in academic writing, such as using AI to generate initial ideas or refine the student's own work, receive somewhat lower mean scores, ranging from 2.35 to 2.75. This suggests that students have a more ambivalent perception of these scenarios, and may not fully understand the potential ethical implications of such uses of AI in academic writing.

Finally, scenarios F8 to F11, which describe the use of AI for more ancillary tasks like checking grammar or searching for resources, are perceived as the least problematic, with mean scores all below 2.35. This suggests that students see these uses of AI as legitimate tools to support their academic writing, rather than as instances of academic misconduct.

In general, these findings suggest that students have a nuanced understanding of the ethical implications of using AI in academic writing, recognizing the potential for misconduct in certain scenarios while also appreciating the utility of AI as a tool to support their academic work. However, the variability in students' perceptions, particularly for the more nuanced scenarios, indicates that there is room for further education and discussion about the appropriate use of AI in academic writing. In fact, without proper policy or guidelines, it is difficult for students and teachers to clearly identify what is AI misconduct.

It is noteworthy that the standard deviations for all items in this survey were over 1 suggests a high degree of variability in the students' responses. In other words, the students' perceptions of what constitutes academic misconduct when using AI-generated content are quite diverse and differ significantly from each other. This can reflect the subjective nature of the issue, with differing individual interpretations of academic integrity and the ethical use of AI.

In this case, it signifies the need for further discourse, clarification, and perhaps educational initiatives to develop a more unified understanding and set of standards for the use of AI in academic work. It also highlights the challenge in establishing clear-cut rules in relation to AI usage in academic settings, as student perceptions are varied and there is no consensus.

## 9. Discussion

The concept of AI-giarism is inherently complex, given that it stands at the intersection of two highly dynamic fields: AI technology and academic integrity. This study set out to investigate students' perceptions of what constitutes academic misconduct when using AI-generated content in higher education. This discussion will reflect on the findings of the study, drawing on the existing literature on plagiarism, AI in education, and academic integrity to contextualize these findings and explore their implications.

Firstly, the study's findings revealed a general understanding among students of traditional forms of plagiarism. The strongest agreement was observed for actions where content was directly copied from a source without proper attribution, indicating a solid understanding of plagiarism in its most basic form. However, the presence of some variability in students' perceptions of different forms of plagiarism, as reflected by the standard deviations, suggests the need for continuous education and clarification on the nuances of academic misconduct. Prior research has shown that discrepancies in understanding plagiarism can be attributed to cultural differences, academic disciplines, and the lack of clear, universal definitions (Gullifer & Tyson, 2014). Hence, a concerted effort by academic institutions and educators to clarify and unify definitions of plagiarism can address this gap and perhaps, interventions are needed.

Secondly, students demonstrated a nuanced understanding of AI-giarism, which is noteworthy considering the novelty of this concept. Students viewed the outright use of AI tools to generate and copy content as a significant instance of academic misconduct, thus extending their understanding of plagiarism to the use of AI. However, the more ambivalent perceptions of scenarios involving the more nuanced uses of AI tools suggest that students struggle with the blurred lines between AI as a tool to support academic writing and as a potential enabler of academic misconduct.

The findings underscore the need for clear and specific guidelines on the use of AI in academic work, consistent with recommendations from Chan (2023). Currently, APA and MLA citation guidelines suggest the citation of AI-generated content by providing detailed information about the model, training data, and the organization responsible for its development. Yet, there are challenges, as noted by McAdoo (2023), that the same AI model or dataset can be used by different researchers to generate different outputs, making it hard to trace the origin of specific content. Therefore, as well as evolving citation guidelines, there is a need for further research and discussion on how to address these challenges in order to maintain academic integrity.

There are also ethical considerations. As noted earlier, some AI-generated content may contain sensitive or personal information, and researchers must ensure this is used responsibly and ethically. This raises questions about the use of AI technology in research and the need for clear ethical guidelines (Mittelstadt et al., 2016).

The variation in student responses may reflect the ongoing debate in the academic community about the ethical use of AI in education. Students' ambivalence could be due to the lack of consensus on what constitutes AI-giarism and the absence of clear guidelines and policies. This finding aligns with the concerns raised by academics about the emergence of AI-giarism (Chan & Tsi, 2023; Salvagno et al., 2023). Therefore, to support academic integrity, educational institutions should consider integrating discussions about the ethical use of AI into their academic integrity policies and practices, as well as providing clear guidelines on how to properly cite AI-generated content, both students and teachers need to be AI-literate (Kong et al, 2021; Long & Magerko, 2020; Ng et al, 2021).

One striking finding from this research is that higher education students still do not fully comprehend traditional plagiarism rules despite their zero-tolerance nature. This raises a significant concern, and perhaps underscores the importance of introducing concepts of academic integrity early in the educational journey, potentially creating cultures of integrity for prevention and intervention (Stephens, 2015). Previous research has demonstrated that formative years are crucial for instilling ethical behaviours and understanding, thus early education could establish a strong foundation for academic integrity (Stephens, 2019).

Answering the first research question, defining violations of academic integrity when using AI-generated content in higher education, proved challenging due to the rapidly evolving nature of generative AI technologies. While it is difficult to set concrete rules, the broad acceptance and integration of AI in education is an inevitable shift needed to prepare students for future societal demands (Chan & Zhao, 2023; Chan & Lee, 2023).

The integration of the AI function, Co-pilot (Posey, 2023), into popular Microsoft Office tools may further complicate the understanding and handling of AI-giarism. With ubiquitous tools like Microsoft Word adopting AI technology, the line between academic misconduct and legitimate use of technological assistance in academic work may become even more blurred.

As we conclude, we invite readers to ponder on several questions provoked by this study:

- Do we embrace technology to ensure we are future ready?
- Would the future soon allow human-machine partnership in academic publication?
- Do we uphold the AI-plagiarism standards, and if so, what are they? Where do we draw the line?

- Do we acknowledge AI tools in each step of our work for transparency and accountability? How many prompts and evidence are needed?
- Can there be unintentional AI-plagiarism or only zero tolerance?
- At what point do humans end and AI begin? Is there a possible partnership?

These questions not only delve into the ethical dimensions of AI use in academic contexts but also probe the future trajectory of academia in an AI-driven world. Our study is a stepping stone in this ongoing exploration and sets the stage for further research into AI-giarism. It is clear that this discussion will continue to evolve alongside advancements in AI technology, underlining the importance of continuous research, dialogue, and adaptation in the face of technological advancements in academia.

## 10. Implications

The implications of this study extend far beyond its direct findings. This study carries several significant implications for the future of academia, policy-making, and the integration of AI technology into educational processes.

**Education and Curriculum Development:** The study reveals that there is a degree of ambiguity and a lack of understanding among students about what constitutes AI-giarism. This points to a pressing need for academic institutions to include AI ethics, and particularly issues surrounding AI-giarism, within their curricula. Such education should ideally start at an early stage to ensure that students are well-prepared and ethically informed as they enter higher education and beyond.

**Questionnaire Instrument as Policy and Educational Guidelines:** Our questionnaire instrument, particularly as encapsulated in Figure 1, carry several implications for educators and policymakers. This study has initiated a first step in conceptualizing AI-giarism by providing a scale that can be used as a guideline for defining what constitutes AI academic misconduct. This scale could be instrumental in assisting educators, who often struggle with comprehending and explaining the complex dimensions of AI-giarism. It can serve as a guide to facilitate understanding and discussions around AI academic misconduct, to design pedagogy and assessment with the adoption of AI, thereby fostering an environment that encourages academic integrity in the age of AI. In fact, the scale demonstrates the complexity of AI-giarism and it may be the time for us to rethink the definition of academic misconduct when it comes to AI and education integration.

**Policy-making and Guidelines:** The study highlights a gap in the existing academic integrity policies, which currently do not adequately cover AI-giarism. The findings from this study, especially the insights gleaned from the student survey, can inform the development of comprehensive guidelines on the ethical use of AI in academic work. Policymakers should consider the nuanced nature of AI-giarism, and guidelines should be flexible enough to accommodate future advancements in AI.

**AI Tool Development and Transparency:** This study draws attention to the need for more transparency in AI tool usage. AI developers and providers could potentially create features that allow users to automatically cite AI-generated content or ideas, thereby promoting ethical usage and reducing unintentional AI-giarism.

This study is a significant step towards understanding the complex issue of AI-giarism. Crucially, the research exposes a clear need for proactive measures within academic institutions to educate students on the ethical implications of AI use in their work. By establishing comprehensive and adaptable policies on AI ethics and academic integrity, we can help prepare students for responsible participation in an increasingly AI-integrated future.

**11. Conclusions**

This study represents a pioneering effort in the exploration of the emerging issue of AI-giarism in the higher education context. The findings have shed light on the complex attitudes and understandings of students in relation to both traditional plagiarism and AI-giarism. Although students display an understanding of the outright unethical uses of AI in academic writing, their perceptions of more nuanced uses highlight a need for further exploration, education and comprehensive policy guidelines.

**Future Research**

The study underscores the need for further research on AI-giarism. This is a rapidly evolving area with many complexities, and continuous research will be needed to keep up with changes and challenges that arise. Future research could explore the perspectives of other stakeholders, such as educators and administrators, and examine the effectiveness of various strategies to prevent AI-giarism. Experimental studies could be conducted to test the effectiveness of various educational interventions or policy changes. Longitudinal studies could track changes in students' attitudes and behaviours over time, yielding insights into the long-term impacts of AI integration into academic work.

Given the global nature of both AI technology and higher education, international collaboration could be beneficial in addressing AI-giarism. Collaborative efforts could lead to the development of globally recognized guidelines, educational initiatives and tools that promote the ethical use of AI in academia worldwide.

**Limitations**

The study has several limitations, including a potential lack of representativeness due to the convenience sampling technique used. A mixed-methods approach could potentially reveal important insights that were not accessible through the survey alone. For example, qualitative data could shed light on why students have more ambiguous attitudes towards the more nuanced uses of AI in academic writing, or why there is such variability in students' responses. These insights could be invaluable for developing effective educational interventions and policy guidelines on AI-giarism. The rapidly evolving nature of AI technology also means that the findings could quickly become outdated. Furthermore, the study's focus on students' perceptions leaves other key perspectives, such as those of educators and administrators, unexplored.

This study provides an initial exploration of students' perceptions of AI-giarism, but it also opens up numerous avenues for further research. As AI technology continues to evolve and becomes more integrated into educational contexts, there is a need for ongoing dialogue, research, and education about AI-giarism. Developing clear and comprehensive guidelines for the ethical use of AI in academic work is an essential step towards maintaining academic integrity in the age of AI.

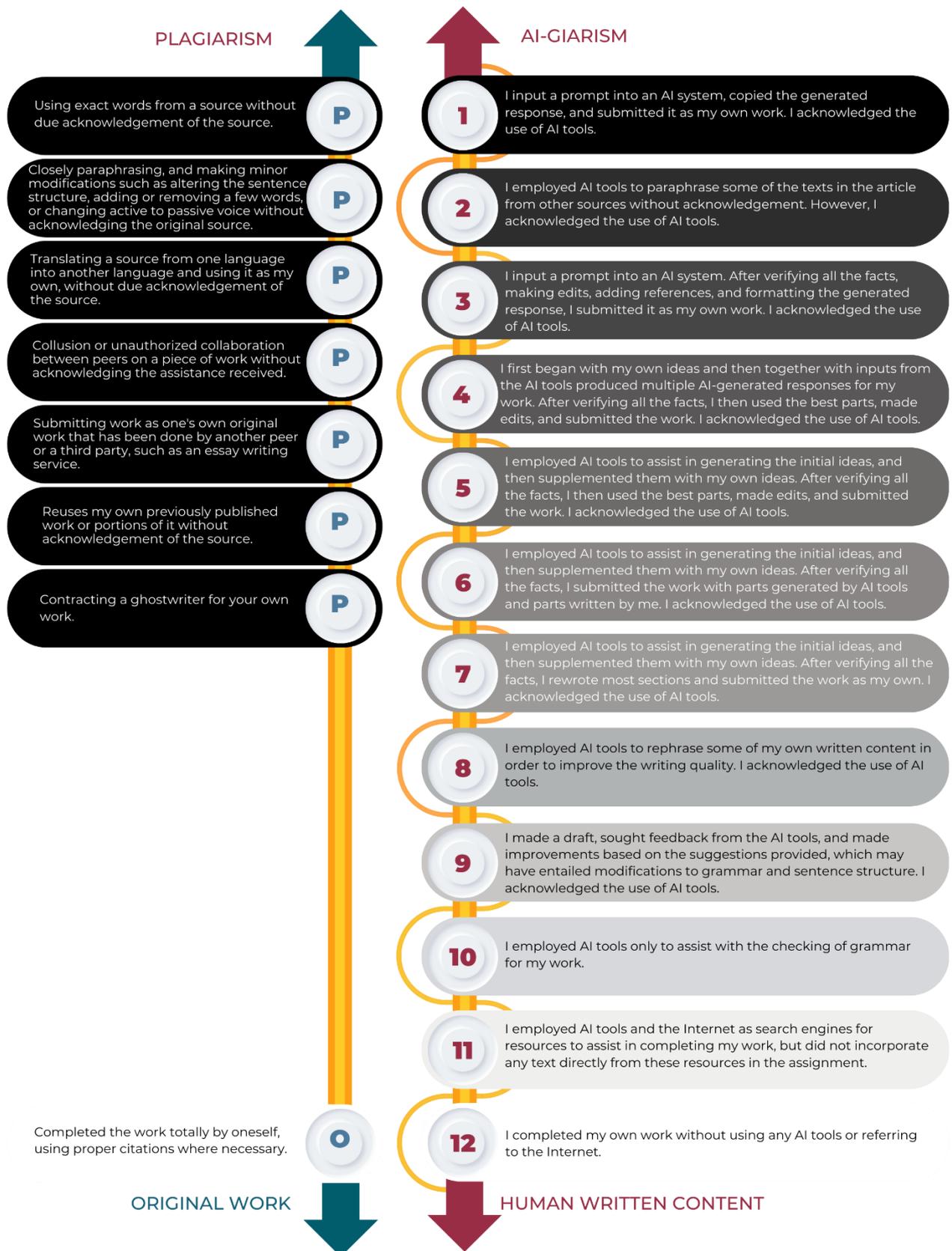

Figure 1: AI-giarism and Plagiarism Literacy for academic writing – what constitutes unethical?

I declare no competing interests.